\documentclass[12pt]{article}
\usepackage{multirow}
\usepackage{amsmath,amsfonts,amssymb,bm,slashed,bbm,mathrsfs}
\usepackage{graphicx,color}
\renewcommand{\Ref}[1]{(\ref{#1})}

\newcommand{\nn}{\nonumber}

\renewcommand{\Ref}[1]{(\ref{#1})}
\newcommand{\beao}{\begin{eqnarray*}}
\newcommand{\eeao}{\end{eqnarray*}}
\newcommand{\be}{\begin{equation}}
\newcommand{\ee}{\end{equation}}
\newcommand{\bea}{\begin{eqnarray}}
\newcommand{\eea}{\end{eqnarray}}
\newcommand{\beq}{\begin{eqnarray}}
\newcommand{\eeq}{\end{eqnarray}}

\begin{document}



\title{Effective color charge in  high temperature  \\ gluon   plasma at  $A_0$ background }

\author{ V. Skalozub\thanks{e-mail: Skalozubvv@gmail.com}\\
{\small Oles Honchar Dnipro National University, 49010 Dnipro, Ukraine}}

\date{\today}
\maketitle\thispagestyle{empty}







\begin{abstract}
In   high temperature $SU(2)$ gluodynamics, in a relativistic background $R^{ext.}_\xi$ gauge, the effective color charge  $\bar{g}^2(A_0) $  is calculated in the presence of the   $A_0$ condensate which is related to  Polyakov's loop and also  serves as the order parameter of the deconfinement phase transition at high temperature. The gauge-invariant value of the condensate $ (A_0^{cl})_0$   is obtained from the two-loop effective potential of order parameter $W_L(A_0^{cl})$ which is independent of  the gauge-fixing parameter $\xi$ and  has a nontrivial minimum position. Due to the smallness of the expansion parameter  $\sim \frac{g^2}{8 \pi^2}$, this potential is applicable at high temperatures beginning from the deconfinement temperature $T_D$.  The   effective charge accounts for a correlation  between the one- and the two-loop terms of the $W_L(A_0^{cl})$.  It results in  decreasing of the effective gluon interactions  in the plasma, compared to the zero condensate case.  Temperature dependence of $\bar{g}^2((A_0^{cl})_0)$  is investigated.  The                   importance of the found dependence  for different applications is noted.

\end{abstract}


\section{Effective  potential  of order parameter}	

In quantum chromodynamics at high temperature, a new matter state - quark-gluon plasma - is created \cite{aoki09-6-088}. Standard information on the deconfinement phase transition is adduced, in particular, in
 \cite{bell96}. The order parameter of the phase transition is the Polyakov loop.
  In $SU(2)$,  it is defined as
\be \label{PL} \left <  L \right>  = \frac{1}{2}  Tr P \exp\bigl ( i g \int_0^\beta d x_4 ~A_0(x_4, \vec{x})\bigr) . \ee
Here, $g$ is coupling constant,  $A_0(x_4, \vec{x})  = A^a_0(x_4, \vec{x} ) \frac{\tau^a}{2}$ is the zero component   of the gauge field
potential, $\tau^a$ is Pauli matrix,  $\beta= \frac{1}{T}$ is inverse temperature and  integration  is going along the fourth direction in  the  Euclidean space-time.

 The  background of the gluon plasma is formed out of so-called $A_0$ condensate which is a constant constituent   of the   gluon field potential $ Q_\mu = [Q_0 = A_0 + (Q_0)_{rad.}, (Q_i)_{rad.}]$ and $ Q_{rad.}$ are radiation fields. This  condensate is correlated to the Polyakov loop and also could serve as the order parameter for the  phase transition. The  $A_0$  order parameter is  a special type  solution of field equations whereas the latter one can be calculated for an arbitrary field configuration. At low temperatures $T < T_d,$  $A_0 = 0$. For $T > T_d , A_0 \not = 0,$ where $ T_d$ is a phase transition temperature. In this phase color is  opened and  colored states are realized.

The generation of $A_0$ has been  derived by different methods in various models from a two-loop effective potential $W(A_0) = W^{(1)}(A_0) + W^{(2)}(A_0)$. Review paper on this topic is \cite{bori95-43-301} where detailed information is presented. The results of calculation for SU(2) gluodynamics in the relativistic  background $R_\xi(A_0)$  gauge  read \cite{bely91-254-153}, \cite{skal92-7-2895}, \cite{skal21-18-738},
\bea \label{EP} W(x) &=& W^{(1)}(x)+ W^{(2)}(x),\\ \nn
\beta^4 W^{(1)}(x)&=&  \frac{2}{3} \pi^2 [ B_4(0) + 2 B_4(\frac{x}{2}) ], \\ \nn
\beta^4 W^{(2)}(x)&=&\frac{1}{2} g^2 [  B_2^2(\frac{x}{2})+ 2 B_2(0)) B_2(\frac{x}{2})] +
\frac{2}{3} g^2(1 - \xi) B_3(\frac{x}{2})B_1(\frac{x}{2}),\eea
where $B_i(x)$ are Bernoulli's polynomials, $x = \frac{g A_0 \beta}{\pi}$.
The Bernoulli's polynomials defined $ modulo$ 1   are
\bea \label{BP} B_1(x)&=&x - \frac{x}{2 |x|}, ~~ B_2(x) = x^2 - |x| + \frac{1}{6}, \\ \nn
B_3(x)&=& x^3 - \frac{3}{2} \frac{x^3}{|x|} + \frac{1}{2} x, \\ \nn
B_4(x)&=& x^4 - 2 |x|^3 + x^2 - \frac{1}{30}. \eea
At $x = 0 $ the $B_1(x)$ is defined to be 0. Because of the symmetry of polynomials, a periodic structure in internal space is
realized (for details see \cite{bori95-43-301}). Bellow, we consider the main interval of  $x, ~0 \leq x \leq 2$.

To find minima of $W(x)$ we apply an expansion in powers of $ g^2$ and get
\bea \label{W0} \beta^4 W_{min}& =&  \beta^4 W (0)
-\frac{1}{192\pi^2}(3-\xi)^2g^4, \nn \\
&&x_0 =  g^2 \frac{(3 - \xi)}{8 \pi^2} ,\eea
where the first term is the value at $x = 0$.
As we see, both the minimum position and the minimum energy value are gauge-fixing dependent. Hence the gauge invariance of the $A_0$ condensation phenomenon is questionable.

 In \cite{skal21-18-738}, to  obtain a gauge-fixing-independent effective potential related to  Polyakov's loop, the following  procedure has been developed. The expression \Ref{EP} was calculated on a special orbit in the $(x, \xi)$-plain where the $\xi$-dependence is cancelled in the total in the order $g^2$,  due to the variation of  $A_0$ which also ensures  expressing  $W(x)$ through   Polyakov's loop.      This relation looks as  (eq.(17) in \cite{skal21-18-738})

\be \label{xclas} x = x_{cl} + \frac{g^2}{4 \pi^2} B_1 (\frac{x_{cl}}{2}) ( \xi + 1) ,\ee
where $x$ is "nonphysical nonrenormalized" value and $x_{cl}$ is observable "physical" one.
This formula follows  from the  one-loop correction to the Polyakov loop.  To obtain $\xi$-independent expression we have to substitute \Ref{xclas} in \Ref{EP} and expand in powers of $g^2$. As a result, $\xi$-dependent parts are mutually cancelled and we find the effective potential of the order parameter $W_L(x_{cl})$.

 Formula \Ref{xclas} identically coincides  with the characteristic line in the $(\xi, x)$-plane coming through the point $\xi = - 1, x_{cl}$. This curve is obtained by the Nielsen identity method for the effective potential \cite{skal21-18-738}. In such a way these two gauge independent approaches are related.  This important correspondence  is also discussed in details  for full QCD with finite chemical potential in \cite{bord21-81-998}.

   A posteriori described procedure can be    realized     if we set in \Ref{EP} $x \to x_{cl}, ~ \xi \to - 1$. The same, of course, refer to the minimum values \Ref{W0}.  They are the gauge invariant characteristics of the background state in the plasma. In particular,  Polyakov's loop in the minimum   equals to
$< L > = \cos( \frac{ g^2}{4 \pi})$.
\section{Effective color charge }
Now we turn to investigation of the influence of $A_0^{cl}$  on the coupling parameter.
To find such  dependence,   we introduce the definition
\be \label{EQ}  \frac{1}{\tilde{g}^2} = \frac{1}{g^2}  \frac{  W}{  W^{(1)}} = \frac{1}{g^2} \Bigl( 1 + \frac{ W^{(2)}(x)}{ W^{(1)}(x)}\Bigr), \ee
where $\tilde{g}^2$ is an effective color charge squared, which is  generated in the $A_0$ presence   due to the correlation of the two-loop, $W^{(2)}(x)$, and the one-loop, $W^{(1)}(x)$,  contributions in \Ref{EP}. Note again that here $ x = (x_{cl})_{0}$ \Ref{W0}, and for $\xi = - 1 , ~~ (x_{cl})_{0} = g^2 \frac{1}{2 \pi^2}$. In one loop order the condensate is absent, and $\tilde{g}^2 = g^2$.

 Formula \Ref{EQ} corresponds to the definition of the effective electric charge $\tilde{e}^2$ in a magnetic field in QED,  or gluon field coupling $\bar{g}^2$ in color magnetic field, at finite temperature \cite{skal96-11-5643}. The main difference is that in both latter cases the influence of vacuum polarization  follows from the correlations of the one-loop and  tree-level terms in  corresponding effective potentials.
That is, we have to replace in \Ref{EQ}   $W^{(1)}(x) \to \frac{H^2}{2} $ and  $W^{(2)}(x) \to W^{(1)}(H). $

 At first glance, this is a simple mathematical difference which does not depend on the way of field generations. In the case of magnetic field, one may assume that it is produced by some external source which   maintains the field of arbitrary strengths. Contrary, in the case of $A_0$ any external source is present basically. This field is a solution to Yang-Mills field equations subjected to the periodicity conditions in the imaginary time. It has no sources. So, the only mechanism for generating this field is a spontaneous condensation happening when higher order contributions are taken into consideration. The definition \Ref{EQ} accounts for this possibility.

Other important point is that the coupling $g^2$ could also   depend on different external parameters - temperature, fields, density,  etc. Relation \Ref{EQ} accounts for the presence of them through the value of $(x_{cl})_{0}$.
 As a result,  we  obtain   the effect of the condensate in the plasma.

Moreover, because the ratio of the one- and two-loop quantum corrections   is proportional to $\frac{g^2}{8 \pi^2}$, we can verify that   the same value of the condensate follows if we take into consideration in Bernoulli's polynomials the terms of the order $g^2 $ in \Ref{EP}. That is $x^2 - \frac{1}{30}$ in $B_4(x)$ and the linear in $x$ terms in the $W^{(2)}$ which is of the order $g^2$. This effective potential is
\be \label{asep} W^{as.} = \frac{2}{3} \pi^2 \Bigl( \frac{x^2}{2} - \frac{1}{30}\Bigr) - \frac{g^2}{12} \Bigl( (3 - \xi ) x -  \frac{1}{6}\Bigr). \ee

To investigate the behavior of $\tilde{g}^2$  as  function of $x_{cl} $, we    insert $ (x_{cl})_{0} = g^2 \frac{1}{2 \pi^2}, \xi = - 1$.  Since the charge $g^2 $ is temperature dependent and (due to asymptotic freedom) it  decreases  with temperature increase, we calculate the effective charge for  series of decreasing values. The results are adduced in Table 1
\begin{table}[ht]
    \caption{Effective charge squared  $\tilde{g}^2$ as function of $g^2$ } \label{tb:Constraints}
    \centering
    \begin{tabular}{|c|c|c|c|c|c|c|c|}
        \hline

        $g^2$ & 10 & 7 & 6 & 5 & 3 & 1 &  0.1 \\ \hline
        $\tilde{g}^2$ & 4.3637 & 4.6670 & 4.5697 & 4.2933 & 3.0517 & 1.0461 & 0.1006 \\ \hline


    \end{tabular}
\end{table}

In first line we give a selected series of the charge squared. In the second one we present the corresponding values $\tilde{g}^2$ calculated according \Ref{EQ}, \Ref{asep}.
As we see,  the effective interactions in the plasma are considerably changed. At high temperatures, the presence of the condensate is not important. But at temperatures close to the deconfinement $T_d$, effective coupling $\tilde{g}^2$ significantly decreases due to the condensate.   Such unexpected behavior should be taken into account when the deconfinement phase transition and the high temperature plasma are investigated.
 In particular, the values of $\tilde{g}^2$ have to be  used in all perturbation  calculations of radiation corrections. It is important that the expansion parameter of the theory  remains small, $\frac{g^2 }{8 \pi^2} \sim \frac{1}{10}$. So, the effective potential \Ref{EP} is the reliable approximation beginning the deconfinement phase transition temperatures.

 Similar  behavior of effective couplings takes also place in other models, in particular, in full QCD. The actual values of
$\bar{g}^2$ have to be computed additionally. They must be used in treating  experiment data obtained at  high energy collisions of heavy ions,  in  problems of the early Universe where the high temperature phases with either different $A_0$ condensates or  temperature dependent magnetic fields  determined  by relevant gauge groups  had existed. These questions we left for the future.

\section*{Acknowledgments}
Author indebted Alexander Pankov and Oleg Teryaev for useful discussions and suggestions.


\end{document}